\documentclass[12pt]{iopart}
\usepackage{epsfig,color}
\usepackage{here}
\usepackage{iopams}
\usepackage{amstext}
\usepackage{graphicx}
\usepackage{graphics}

 \def\be{\begin{equation}}
 \def\ee{\end{equation}}
 \def\bea{\begin{eqnarray}}
 \def\eea{\end{eqnarray}}
 \def\gsim{\mathrel{\rlap{\lower0.2em\hbox{$\sim$}}\raise0.2em\hbox{$>$}}}
 \def\ksim{\mathrel{\rlap{\lower0.2em\hbox{$\sim$}}\raise0.2em\hbox{$<$}}}
 \def\kg{\mathrel{\rlap{\lower0.25em\hbox{$>$}}\raise0.25em\hbox{$<$}}}

\begin{document}

\title{Competition of Heavy Quark Radiative and Collisional Energy Loss in Deconfined Matter}

\author{P.B. Gossiaux, J. Aichelin, T. Gousset and V. Guiho}

\address{SUBATECH, Universit\'e de Nantes, EMN, IN2P3/CNRS
\\ 4 rue Alfred Kastler, 44307 Nantes cedex 3, France}
\ead{pol.gossiaux@subatech.in2p3.fr}
\begin{abstract}
We extend our recently advanced model on collisional energy loss of heavy quarks
in a quark gluon plasma (QGP) by including radiative energy loss. We discuss the approach 
and present first preliminary results, including a comparison of the role of both types of 
energy loss for experimental data. We draw the conclusion that the present nuclear 
modification factor $R_{AA}$ data on non-photonic single electrons does not permit 
to ``select'' between this two types of energy-loss mechanisms.
\end{abstract}

%\maketitle

\section{Introduction}
One of the main objectives of the studies of ultrarelativistic heavy ion collisions in heavy 
ion collisions is the exploration of the properties of a quark gluon plasma (QGP), 
which is most probably 
% - according to the prediction of lattice gauge calculations - will be 
formed in these collision. To achieve this objective is all but easy because this plasma,
if created at all, will live only for a very short time before it suffers a phase transition
(or more precisely a cross over) towards the hadronic phase. After this transition the 
hadrons continue to interact weakening the possible information on the plasma phase.

Indeed, the multiplicity of light particles, those which contain only up, down and strange 
quarks, is well reproduced in statistical models indicating that these multiplicities are 
established at temperatures close to the transition temperature between the plasma and the 
hadronic phase and carry therefore little or no information on the plasma interior. The 
multiplicity of the observed resonances presents evidence that, after the multiplicity is 
established, hadronic interactions modify considerably the momenta of the hadrons. In view 
of the fact that many of the  involved elementary cross section are little known or unknown 
it is difficult if not impossible to asses the spectra of the hadrons at the moment of their
formation, a prerequisite for an analysis of the plasma properties with help of these
particles.

The observation that the bulk part of the observed particles is not sensitive to the QGP 
properties limits severely the possibilities to study the interior of the QGP and it is 
all but easy to identify those observables which may serve for this purpose. Those which 
have been advanced include the high $p_T$ hadrons which originate from jets which do not 
come to equilibrium with the plasma as well as the $p_T$ and $v_2$ distribution of heavy 
mesons which contain either a $c$ or a $b$ quark.

Heavy quarks are produced in hard binary initial collisions between the incoming nucleons. 
Their production cross sections are known from pp collisions and their spectra can be 
calculated (see f.i. \cite{Cacciari:2005rk}) in the framework of perturbative Quantum 
ChromoDynamics (pQCD), up to a systematic error band due to the choice of various scales. 
In heavy ion collisions, these spectra can be taken bona fide as the initial transverse momentum 
distribution of heavy quarks. Comparing this distribution with that measured allows to define 
the nuclear modification factor 
$R_{AA}=(d\sigma_{AA}/d^2p_T)\ / \ (N_c\, d\sigma_{pp}/d^2p_T)$, where $N_c$ is the number 
of initial binary collisions between projectile and target. The deviation of $R_{AA}$ from 
unity essentially reflects the interaction of the heavy quark with the plasma because the 
hadronic cross sections of heavy mesons are small. Present data shows that ``high'' $p_T$ 
heavy quarks ($p_T\gtrsim m$) do not come to thermal equilibrium with the QGP; 
therefore $R_{AA}$ contains the information on the interaction of heavy quarks while they 
traverses the plasma\footnote{Although one observes similar fact for light-quark leading 
hadrons, it is not clear whether the dominant basic microscopic processes are of similar 
nature for these two probes, due to the heavy quark finite-mass and the flavor exchange 
mechanism.}. In addition, the distribution of heavy quarks at the moment of their creation 
is isotropic in azimuthal direction, therefore the elliptic flow
$v_2 = <\cos2(\phi- \phi_R)>$, where $\phi$ ($\phi_R $) is the azimuthal angle of the 
emitted particle (reaction plane) is 0. The observed  finite $v_2$ value of the observed 
heavy meson can only originate from interactions between light QGP constituents and the 
heavy quarks. The simultaneous description of $R_{AA}$ and $v_2$ -- presently the only 
observables for which data exist -- gives then the hope to better understand the 
interaction of those heavy quarks with the QGP and thus its profound nature.
Unfortunately the experimental results depend not only on the elementary interaction but 
also on the description of the expansion of the QGP. Therefore the ultimate aim is to 
control the expansion by results on the light meson sector. This has not been achieved yet 
and therefore it is difficult to asses the influence of the expansion on the observables.

Coming to facts, the $R_{AA}$ of $\approx 0.2-0.3$ value observed for large $p_T$ 
($p_T\sim 4-8\,{\rm GeV}$) non-photonic single-electrons (n.p.s.e.)
\cite{Abelev:07,Adare:06b}, of the same order of the pionic $R_{AA}$, is associated 
with 2 related puzzles, hereafter referred as ``single electron puzzle'' (s.e.p.):
\begin{itemize}
\item level-1 s.e.p.: One cannot understand such a small value within the framework of pQCD;
\item level-2 s.e.p.: Even if one introduces some degree of freedom in the pQCD inspired models,
i.e. some free parameter as f.i. $\hat{q}$ or $dN_g/dy$, it is not possible to reproduce both 
$R_{AA}$ of n.p.s.e. and pions within the same global framework.
\end{itemize}

To substantiate level-1 s.e.p., we notice that the elastic cross section and hence the collisional energy loss 
introduced in early approaches \cite{Svetitsky:1987gq} has to be multiplied \cite{Gossiaux:2004sqm} by an 
artificial $K$ factor of the order of 10 to match the experimental data. 
These early calculations, however, used ad hoc assumptions on the (fixed) coupling constant $\alpha_s$ and 
on the infrared regulator $\mu$. This last drawback was partly cured in \cite{Moore:2004tg}, where an HTL
regularization is used for the collisional energy loss, but where the spatial diffusion coefficient $D_s$ one 
should privilege to reproduce the experimental data is still found much smaller then its pQCD value\footnote{One
should moreover notice that no $b$-quark contribution to n.p.s.e. was considered in 
\cite{Gossiaux:2004sqm,Moore:2004tg}, which is known to increase the $R_{AA}$.}. Of course, collisional energy loss 
is only one source of the energy loss. For light particle the radiative energy loss is even more important.
Its importance for heavy quarks has been addressed in numerous publications
(\cite{Djordjevic:04a,Djordjevic:04b,Armesto:04,Djordjevic:2005a,Djordjevic:2006a,
Armesto:06,Wicks:07} and references 
therein, to cite but a few) and is however still debated. 
In \cite{Djordjevic:2005a,Djordjevic:2006a} and in previous 
works of the authors, the heavy quark quenching was {\em predicted} to be undeniably larger than later observed in 
experiment. This work was pursued in \cite{Wicks:07}, where both radiative {\em and} collisional processes are 
implemented, as well as path length fluctuations (not considered before by the authors)\footnote{but however with 
a rather simple implementation of medium evolution (no radial expansion, medium considered at a fixed Bjorken time)}. 
% probability of coll Eloss should not be taken as Gaussian; misses the flavor exchange mechanisms
With the parameters chosen -- fixed $\alpha_s=0.3$ and $dN_g/dy=1000$ -- and both additional mechanisms \cite{Wicks:07} 
could maintain the good agreement found for the pion $R_{AA}$ \cite{Vitev:2002} in the GLV approach with pure radiative 
energy loss and fixed 
path length, while the n.p.s.e. $R_{AA}$ was reduced w.r.t. \cite{Djordjevic:2005a,Djordjevic:2006a} but still found to 
exceed the experimental values\ldots a conclusion that supports the so-called level-2 s.e.p.. In \cite{Armesto:04}, 
Armesto et al. have extended the Wiedemann-Salgado path-integral formalism for radiative energy 
loss~\cite{Wiedemann:2000} to the case of finite-mass 
parton. Later on, they applied their formalism \cite{Armesto:06} to study the n.p.s.e. quenching in the case of 
AA collisions, under the same assumptions used in a previous work dedicated to the light hadrons 
\cite{Dainese:2005} (from which they have extracted $\hat{q}\in[4,14]\,{\rm GeV}^2/{\rm fm}$). Although
Armesto et al. argue that ``claims of inconsistency between theory and experiment are not supported'', they 
obtain a n.p.s.e. $R_{AA}\in[0.4,0.5]$ for $p_T=5-8$ GeV/c, i.e. not enough quenching from their theoretical 
``prediction'' (once calibrated on light hadron quenching).

These puzzles have casted doubt, whether pQCD is the right framework to describe these interactions and 
quenching at $p_T=5-20\,{\rm GeV}/c$, and alternative approaches were proposed, based on the advocated existence of 
non-perturbative degrees of freedom in the QGP~\cite{VHR:05} or on the AdS/CFT conjecture \cite{Horowitz:2008,
Horowitz:2008b}. 
In a more conservative way, Peshier \cite{Peshier:06} has argued that the collisional energy loss could be vastly
increased if a running coupling is taken to evaluate the elastic cross section. All alternative approaches have 
the common property to hold larger stopping power for heavy quarks than pQCD; the last proposal has however 
the appealing feature that collisional energy loss presents a weaker dependence on the parton mass (than the radiative 
one), what seems to be favored by the level-2 s.e.p.. Inspired by \cite{Peshier:06}, we have recently advanced an 
approach for the collisional energy loss of heavy quarks in the QGP 
\cite{Gossiaux:2008jv,Gossiaux:2009mk,Gossiaux:2009hr} in which a) $\mu$ has been fixed by the demand that more 
realistic calculations using the hard thermal loop approach give the 
same energy loss as the 
Born type pQCD calculation and b) the coupling constant is running and constrained by the sum rule advanced by 
Dokshitzer~\cite{Dokshitzer:02}. Both these improvements increased the cross section especially for small momentum 
transfers and reduced therefore the necessary $K$ (used in our previous work \cite{Gossiaux:2004sqm}) factor to 2.

Faced with the theoretical uncertainties affecting $R_{AA}$, a current has developed \cite{Phenix:08} which
suggests that the ``unknown'' parameters that basically encode the interaction frequency with the medium (like 
$\hat{q}$ and $dN_g/dy$) should be constrained using experimental data. It was even argued by Lacey et 
al.~\cite{Lacey:09} that a global analysis of pions and 
n.p.s.e. $R_{AA}$ was ``consistent with jet quenching dominated by {\em radiative} energy loss for both heavy and 
light partons''. However, scenario's relying on a finite contribution of collisional energy loss were not considered
in \cite{Lacey:09}. One of the goals of this contribution is to complement the work of Lacey et al. by considering 
models containing both types of energy loss and the ``competition'' between them in order to reproduce the data. 
To study this question, we have incorporated a radiative energy-loss mechanism in our 
model~\cite{Gossiaux:2008jv,Gossiaux:2009mk,Gossiaux:2009hr}, not as sophisticated
as those cited earlier but which -- to our belief -- incorporates the dominant aspects in the case of heavy 
quarks. We will describe the model in section \ref{section:model} and we report and discuss our first results in
section \ref{section:Results}.

\section{The Model}
\label{section:model}
\begin{figure}
\begin{center}
\epsfig{file=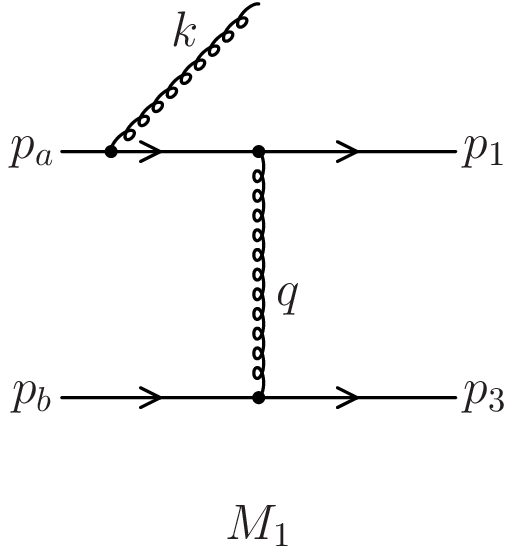,width=0.3\textwidth}
\epsfig{file=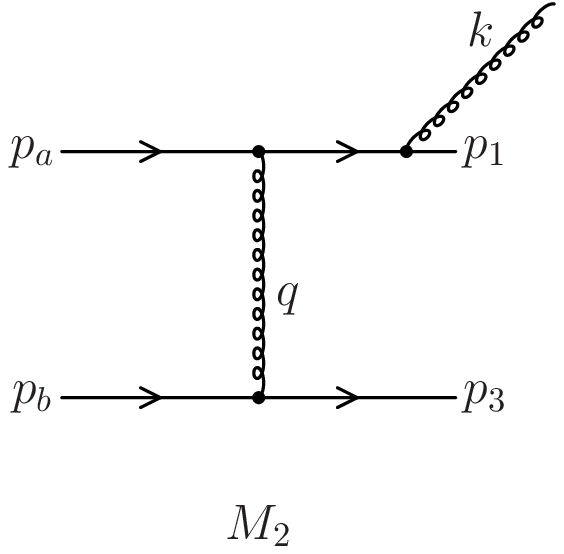,width=0.3\textwidth}
\epsfig{file=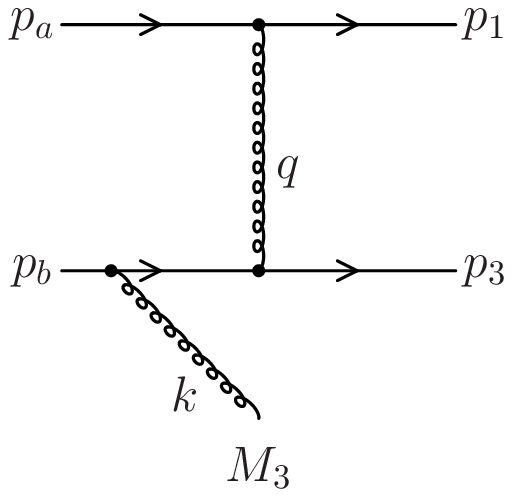,width=0.3\textwidth}
\epsfig{file=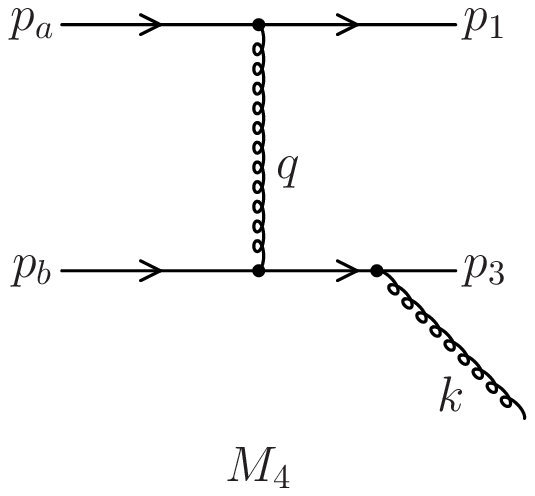,width=0.3\textwidth}
\epsfig{file=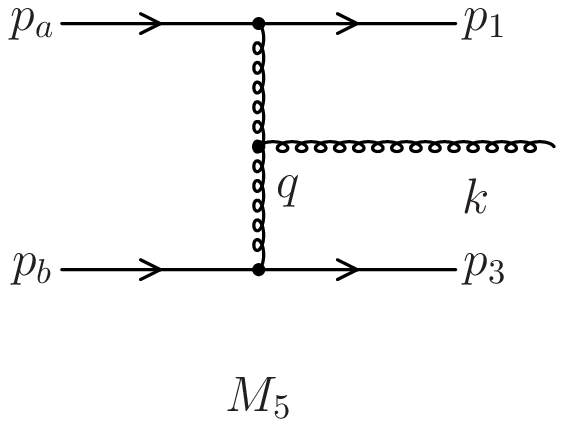,width=0.3\textwidth}
\end{center}
\caption{(Color online) The five matrix elements which contribute to the gluon 
bremsstrahlung in a single process.}
\label{dia}
\end{figure}

Extending our approach to radiative energy loss, we will first focus on gluon emission
in single high energy processes, as described in QCD by the 5 
matrix elements depicted in fig. \ref{dia}.
The commutation relation $T^g T^h=T^hT^g-if_{ghc}T^c\label{tm}$
allows to regroup the 5 matrix elements into
3 combinations, each of them being independently gauge invariant:
\begin{equation} 
\hspace{-1.5cm}
iM^{QED}_{h.q./l.q.}= C_{I/I'} i(M_{1/3}+M_{2/4})\quad{\rm and}\quad
iM^{QCD}= C_V i(M_1+M_3+M_5)\,,
\end{equation}
where h.q. (l.q.) marks the emission of the gluon from the heavy (light) quark line. 
$C_I$, $C_{I'}$ and $C_V$ are the associated color algebra matrix elements. The matrix
elements labeled as ``QED'' are the bremsstrahlung diagrams already observed in Quantum 
Electrodynamics (QED), whereas that labeled ``QCD'' is the diagram specific to QCD. 
The QCD diagram is the main object of interest here, as it can be shown to dominate the 
energy loss of heavy quark.

\begin{figure}[H]
\begin{center}
\epsfig{file=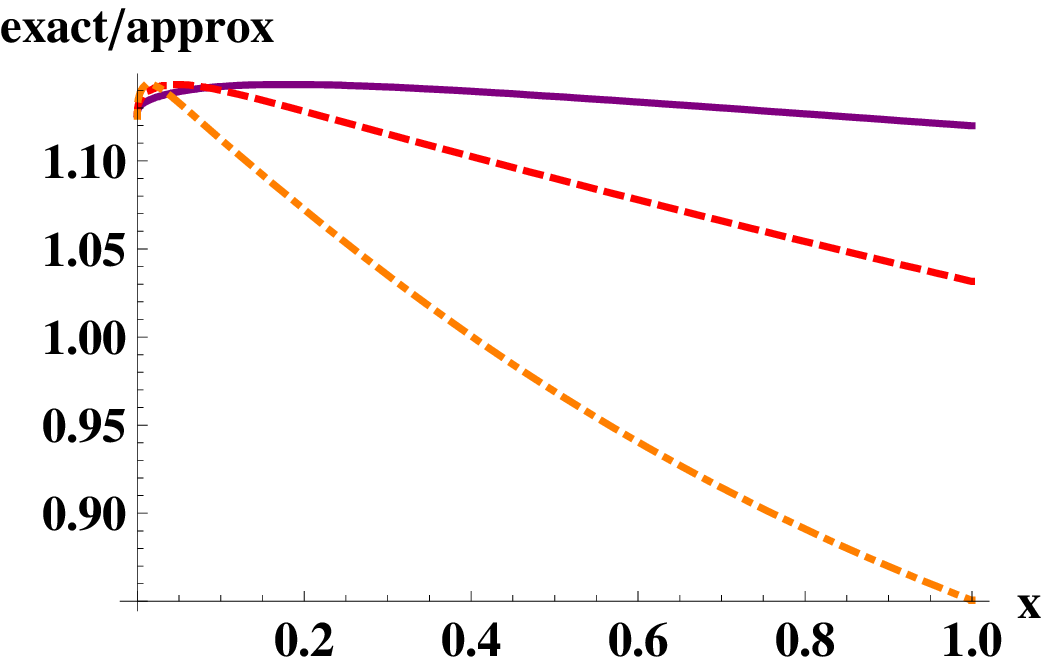,width=0.4\textwidth}
\hspace{1cm}
\epsfig{file=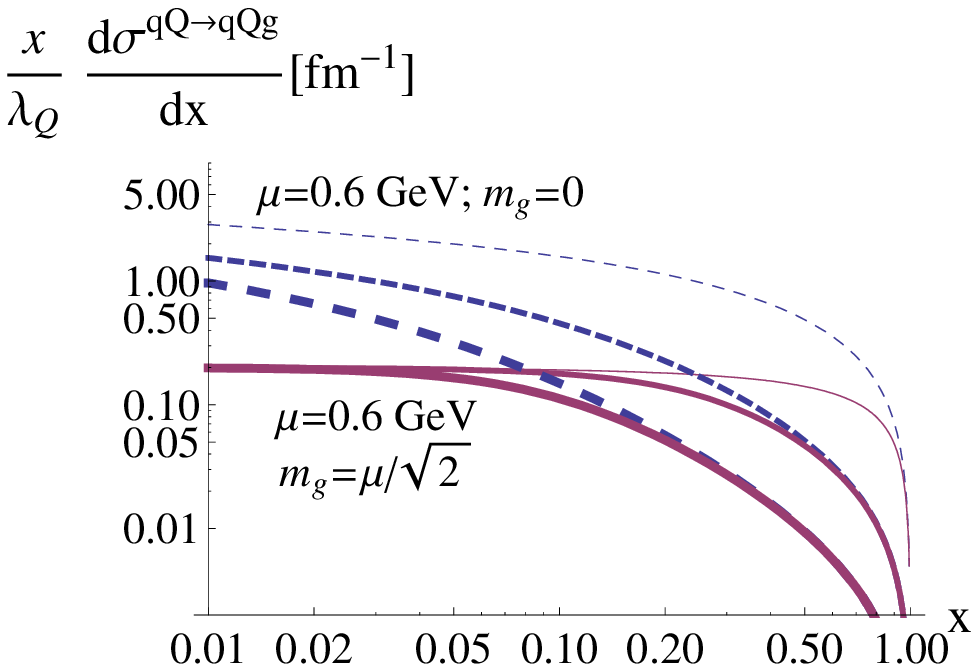,width=0.4\textwidth}
\caption{(Color online) Left: ratio of the the approximate radiation (eq. \ref{polelo}) at 
$\mu = 0.3$ GeV and of the exact $(\vec{q}_T,\vec{k}_T)$ integration of 
(\ref{eq_def_cross_section}) when $s\rightarrow +\infty$ for radiation from strange (full), charm 
(dashed) and bottom (dot-dashed) quarks. Right: radiative energy spectra per unit length with 
(continuous) and without (dashed) gluon mass; thin, plain and thick lines correspond to $m=0.15$, 
1.5 and 5 GeV.}
\end{center}
\label{rat}
\end{figure}

We evaluate the matrix elements in the so-called ``scalar QCD'' approximation, which 
shows to be appropriate at small or moderate gluon fractional momentum $x$. The matrix 
elements in scalar QCD (see ref.\cite{Meggiolaro:1995cu}) are given by
\bea 
\hspace{-1.5cm}iM_1^{\rm SQCD} &=& C_I(ig)^3
\frac{(p_b+p_3)^\mu}{(p_3-p_b)^2}D_{\mu\nu}[p_3-p_b]
\left[\frac{(p_a+p_1-k)^\nu
(2p_a-k)\cdot\epsilon}{(p_a-k)^2-m^2} - \epsilon^\nu \right]
\nonumber\\
\hspace{-1.5cm}iM_5^{\rm SQCD}&=&C_V(ig)^3
D^{\mu\mu'}[p_3-p_b]D^{\nu\nu'}[p_1-p_a]\left[g_{\mu'\nu'}(p_a-p_1+p_3-p_b)_\sigma+
\right.\nonumber\\
&& \left. g_{\nu'\sigma}(p_1-p_a-k)_{\mu'}+g_{\sigma\mu'}(p_b-p_3+k)_{\nu'}\right]
\epsilon^\sigma\times  
%\nonumber \\&&
\frac{(p_3+p_b)^\mu(p_a+p_1)^\nu} {(p_3-p_b)^2(p_1-p_a)^2}\,.
\eea
$M_3$ is obtained by replacing $p_a \to p_b$ and $p_1\to p_3$ in $M_1$. Using light-cone gauge 
and keeping only the leading term in $\sqrt{s}$ we see that the square of the matrix element 
factorizes
\be
|M|^2 = |M_{\rm el}(s,t)|^2 P_g(m,t,\vec{k}_T,x)
\ee
with  $|M_{\rm el}(s,t)|^2 = g^4\frac{4s^2}{t^2}$  being the matrix element squared for the 
elastic cross section in a fixed $\alpha_s$ Coulomb-like interaction.
$P_g(m,t,s,\vec{k}_T)$ describes the distribution function of the produced gluons. 
To discuss the physics we adopt the following light-cone coordinates ($\{p^+,p^-,\vec{p}\}$, 
with scalar product defined as $p_a p_b= \frac{p_a^+p_b^-+p_a^-p_b^+}{2}-p_{a,T}p_{b,T}$):
\begin{equation}
p_a=\{\sqrt{s-m^2},\frac{m^2}{\sqrt{s-m^2}},\vec{0}\}\,,\quad
p_b=\{0,\sqrt{s-m^2},\vec{0}\}
\end{equation}
for the entrance channel and
\begin{equation}
\hspace{-2cm}
k=\{k^+,0,\vec{k}_T\}\,,\;
p_1=p_a+q-k=\{p_1^+,\frac{m^2_T}{p_1^+},\vec{k}_T-\vec{q}_T\}\,,\;
p_3=p_b-q=\{\frac{q_T^2}{p_3^-},p_3^-,-\vec{q}_T\}
\end{equation}
for the exit channel, with $m_T^2=m^2+(\vec{k}_T-\vec{q}_T)^2$,
$p_3^-\approx p_b^--\frac{k_T^2+x m^2_T/\bar{x}}{k^+}$
and $p_1^+\approx \bar{x}p_a^+ -\frac{q_T^2}{p_3^-}$, where $\bar{x}\triangleq 1-x$. In this 
coordinate system, the invariant momentum transfer is $t=q^2 \approx q_T^2$ while
$x$ is defined as $x=k^+/p_a^+$ and represents the relative longitudinal momentum fraction of 
the gluon  with respect to the incoming heavy quark; $|M^{\rm SQCD}|^2$ has moreover a very 
simple form:
\be
\hspace{-0.5cm}|M^{\rm SQCD}|^2= g^2 D^{\rm QCD} 4\bar{x}^2 |M_{\rm el}|^2\left(\frac{\vec
k_T}{k_T^2+x^2m^2}-\frac{\vec k_T-\vec q_T}{(\vec q_T - \vec
k_T)^2 +x^2m^2}\right)^2
\label{qcdem}
\ee
with the color factor $D^{\rm QCD}=C_A*C_{\rm el}^{qq}=\frac{2}{3}$.
The first term in the bracket describes the emission from the incoming heavy quark line, 
the second term the emission from the gluon line. This shows that in light cone gauge and 
in this coordinate system in leading order of $\sqrt{s}$ the matrix element for the emission 
from the light quark do not contribute. In the case of massless quarks we recover the matrix 
elements of Gunion and Bertsch (GB) of \cite{Gunion:1981qs}. From this factorized form, one 
deduces a similar expression for the radiative differential SQCD cross section
when $s\gg m^2$:
\begin{equation}
\hspace{-1.5cm}x\frac{d\sigma^{qQ\rightarrow qQg}}{d^2q_T dx d^2k_T}\approx
\frac{d\sigma^{qQ\rightarrow qQ}_{\rm el}}{d^2q_T}\,P^{\rm SQCD}(m,\vec{q}_T,\vec{k}_T,x)\,  
\Theta\left((m_T +q_T)^2 + \frac{\bar{x}}{x} k_T^2 - \bar{x} s\right)\,,
\label{eq_def_cross_section}
\end{equation}
where 
\begin{equation}
\frac{d\sigma^{qQ\rightarrow qQ}_{\rm el}}{d^2q_T}=\frac{2}{9}\times\frac{4\alpha_s^2}{t^2}\,,
\label{eq_def_el_cross}
\end{equation}
and where the Heaviside function traduces the phase-space boundary which translates into
$k_T\le k_{t,{\rm max}}(\phi(\vec{q}_T,\vec{k}_T))$, 
with $k_{t,{\rm max}}^2\approx x\left[(1-x)s-(\sqrt{m^2+q_T^2}+q_T)^2
\right]$ at small and moderate $x$. Apart from a small region located close to the 
phase space boundary, $P^{\rm SQCD}$ admits the simple form:
\begin{equation}
P^{\rm SQCD}(m,\vec{q}_T,\vec{k}_T,x)\approx\frac{C_A \alpha_s \bar{x}}{\pi^2}
\underbrace{\left(\frac{\vec k_T}{k_T^2+x^2m^2}-\frac{\vec k_T-\vec q_T}{(\vec q_T - \vec
k_T)^2 +x^2 m^2}\right)^2}_{\cal A}\,,
\label{eq_def_P}
\end{equation}
that recovers the GB result at small $x$ when $m$ is taken to 0:
\be
P^{\rm SQCD}(m=0,\vec{q}_T,\vec{k}_T,x\ll 1)= \frac{C_A \alpha_s}{\pi^2}
\frac{q_T^2}{k_T^2(\vec{k}_T-\vec{q}_T)^2}
\label{gbia}.
\ee
We now proceed along the following strategy: in our Monte Carlo numerical framework 
MC$@_s$HQ (results presented in section \ref{section:Results}), we have sampled the 
probability distribution associated to (\ref{eq_def_cross_section}) using
(\ref{eq_def_P}), the collisional cross section described in \cite{Gossiaux:2008jv} 
instead of (\ref{eq_def_el_cross}), and a {\em rigorous implementation of the phase-space 
constrain}. In the rest of this section, we will aim at establishing some synthetic relations
useful for the physical interpretation; for this purpose, we will {\em neglect the phase-space
constrain} in (\ref{eq_def_cross_section}). The integral over $d\phi(\vec{k}_T,\vec{q}_T)$ can 
thus be done analytically pretty easily by rewriting
${\cal A}=\frac{q_T^2+2 x^2 m^2}{e\left((\vec{k}_T-\vec{q}_T)^2+x^2 m^2\right)}
- \frac{x^2 m^2}{e^2}-\frac{x^2 m^2}{\left((\vec{k}_T-\vec{q}_T)^2+x^2 m^2\right)^2}$ with 
$e\triangleq k_T^2+x^2 m^2$. One gets
\begin{eqnarray}
\hspace{-2cm}P^{\rm SQCD}(m,q_T,k_T,x)&\triangleq&
\int d\phi P^{\rm SQCD}(m,\vec{q}_T,\vec{k}_T,x)\frac{2 N_c \alpha_s}{\pi}(1-x)
\nonumber\\ &=& \left[
\frac{q_T^2+2 x^2 m^2}{d\,e}
- \frac{x^2 m^2}{e^2}-
\frac{x^2 m^2 (k_T^2+q_T^2+x^2m^2)}{d^3}\right]
\label{diff_cross_kt}
\end{eqnarray}
with $d\triangleq\sqrt{(k_T^2-q_T^2+x^2m^2)^2+4 x^2m^2 q_T^2}$. 
For $q_T\gtrsim x m$, one observes a dip of the radiation at $k_T\lesssim x m$ as well as 
suppression as compared to light quarks (as can be seen from instance from the value of 
$P^{\rm SQCD}$ at $k_T=0$, $\propto \frac{q_T^2}{(q_T^2+x^2m^2)^2}$). This is the celebrated 
``dead - cone'' phenomenon \cite{Dokshitzer:2001zm}. For smaller $q_T$ values ($q_T\lesssim 
x m$), both cones in (\ref{qcdem}) interfere and the dead-cone structure
disappears, although the radiation of heavy quark stays suppressed. This effect is not 
identified in \cite{Dokshitzer:2001zm} -- where the radiation from the heavy quark is assumed 
to be independent of its diffusion angle -- but indeed corresponds to the most frequent 
situation encountered in induced radiative energy loss as 
$\frac{d\sigma^{qQ\rightarrow qQ}_{\rm el}}{d^2q_T}$ is peaked at small $q_T$.
After $k_T$ integration of $P^{SQCD}(m,q_T,k_T,x)$, one obtains a simple formula which can 
be further approximated by
\begin{equation}
\hspace{-1cm} x\frac{d\sigma^{qQ\rightarrow qQg}}{d^2q_T dx}\slash
\frac{d\sigma^{qQ\rightarrow qQ}_{\rm el}}{d^2q_T}
\approx
P^{\rm SQCD}(m,q_T,x)\triangleq
\frac{2 N_c \alpha_s (1-x)}{\pi}\,\ln\left(1+\frac{q_T^2}{3 x^2 m^2}\right)\,.
\label{eq_def_cross_section2}
\label{eq_def_Q_qt_x}
\end{equation}
In fact, the limitation imposed by the phase space boundary becomes drastic only when
$k_{t,{\rm max}}$ drops below ${\rm max}(x m,q_T)$, i.e. for $x\lesssim q^2_T/s\sim 
\frac{\mu}{E}$, what justifies (\ref{eq_def_Q_qt_x}) in the energy range in which we are 
mostly interested. From (\ref{eq_def_cross_section2}), one sees that the radiative factor 
$P^{\rm SQCD}$ partly cures the IR divergence of the elastic cross section at small $q_T$ 
but it still suffers from a logarithmic divergence. As we consider the radiation of heavy 
quarks propagating through the QGP, we introduce an infrared regulator $\mu^2$ of the order 
of the Debye mass $m_D^2$ in (\ref{eq_def_el_cross}) and can proceed to the $\vec{q}_T$ 
integration, assuming $\mu^2 \ll m^2$ :
\be
x\frac{d\sigma^{qQ\rightarrow qQg}}{dx}=\frac{4C_F\alpha_s^3}{3}\left\{
\begin{array}{ll}
\frac{\ln\left(\frac{\mu^2}{3 x^2 m^2}\right)}{\mu^2}& \text{for $x\lesssim \frac{\mu}{m}$}\\
\frac{\ln(\frac{3x^2m^2}{\mu^2})}{x^2m^2} & \text{for $x\gtrsim \frac{\mu}{m}$}
\end{array}\right.\,,
\label{polelo}
\ee
which describes the radiation spectra in the mass range of interest better than 15\%, as seen 
in fig. \ref{rat}.

In the QGP environment the radiated gluons polarize the medium, an effect that can be 
incorporated phenomenologically to the formalism \cite{Djordjevic:04a} by imposing 
a thermal gluon mass $m_g$ of the order of the Debye mass. In this case the previous 
calculations still hold if $x^2 m^2$ is replaced by $x^2 m^2+(1-x) m_g^2$. On fig. \ref{rat} 
(right) we illustrate the effect of a finite gluon mass on the energy spectra per unit length, 
defined as $\frac{x}{\lambda_Q}\times \frac{d\sigma^{qQ\rightarrow qQg}}{dx}$, where
$\lambda_Q$ is the mean free path of the heavy quark w.r.t. $\sigma_{\rm el}$.
In the region of large $x$ which dominates the {\em average} energy loss, the effect of the 
gluon mass on the radiative cross section of heavy quark is moderate, but a finite gluon mass 
has a drastic effect on the radiation for small and intermediate $x$-values which are usually 
thought to dominate the {\em quenching}~\cite{Baier:01} observed in $R_{AA}$, i.e. $x\lesssim 0.2$. 

To implement the radiative processes in MC$@_s$HQ \cite{Gossiaux:2008jv}, 
-- designed to simulate the heavy quark transport in an expanding medium -- we have
evaluated the total number of radiated gluon candidates in a given elastic collision of 
momentum transfer $q_T$ as\footnote{the $x_{\rm min}$ cut off, chosen as 
$x_{\rm min}=0.05$, permits to discard ultrasoft gluons that lead to IR catastrophe 
due to the ``$x^{-1}$'' weighting of the cross section but which contribute little to the
stopping.} $P^{\rm SQCD}(m,q_T)\triangleq\int_{x_{\rm min}}^{1} \frac{dx}{x} P^{\rm SQCD}(m,q_T,x)$, 
with $P^{\rm SQCD}(m,q_T,x)$ defined by (\ref{eq_def_Q_qt_x}), as well as the average number 
of radiated gluon ensuing an elastic collision as 
$P^{\rm SQCD}(s)=\frac{1}{\sigma_{\rm el}} \int d^2q_T \frac{d\sigma_{\rm el}}{d^2q_T}
P^{\rm SQCD}(m,q_T)$, where the elastic cross section has been taken according with an effective 
running coupling (model ``E'' of \cite{Gossiaux:2008jv}), although the gluon emission embedded 
in $P(m,q_T)$ has been taken at a fixed $\alpha_s$ value in this preliminary study. For each 
elastic collision happening in the MC evolution, we then generate the gluon candidates according 
to Poisson statistics with an average of $P^{\rm SQCD}(s)$, which happens to be $<1$ for 
reasonable values of $\alpha_s$. We then successively sample the 
$\frac{d\sigma_{\rm el}}{d^2q_T}P^{\rm SQCD}(m,q_T)$, $P^{\rm SQCD}(m,q_T,x)$, 
$P^{\rm SQCD}(m,q_T,k_T,x)$ and $P^{\rm SQCD}(m,\vec{q}_T,\vec{k}_T,x)$ weights for $q_T$, $x$, $k_T$ 
and $\phi(\vec{k}_T,\vec{q_T})$, so generating the full kinematics of the $2\rightarrow 3$ process,
and only accept the event if phase-space constrain (\ref{eq_def_cross_section}) is satisfied. A similar 
method is used for the $gQ\rightarrow gQg'$ case. The thermal gluon mass was taken as $m_g = 2T$ at 
each space-time point of the QGP evolution.

\section{Results and Discussion}
\label{section:Results}
\begin{figure}
\begin{center}
\epsfig{file=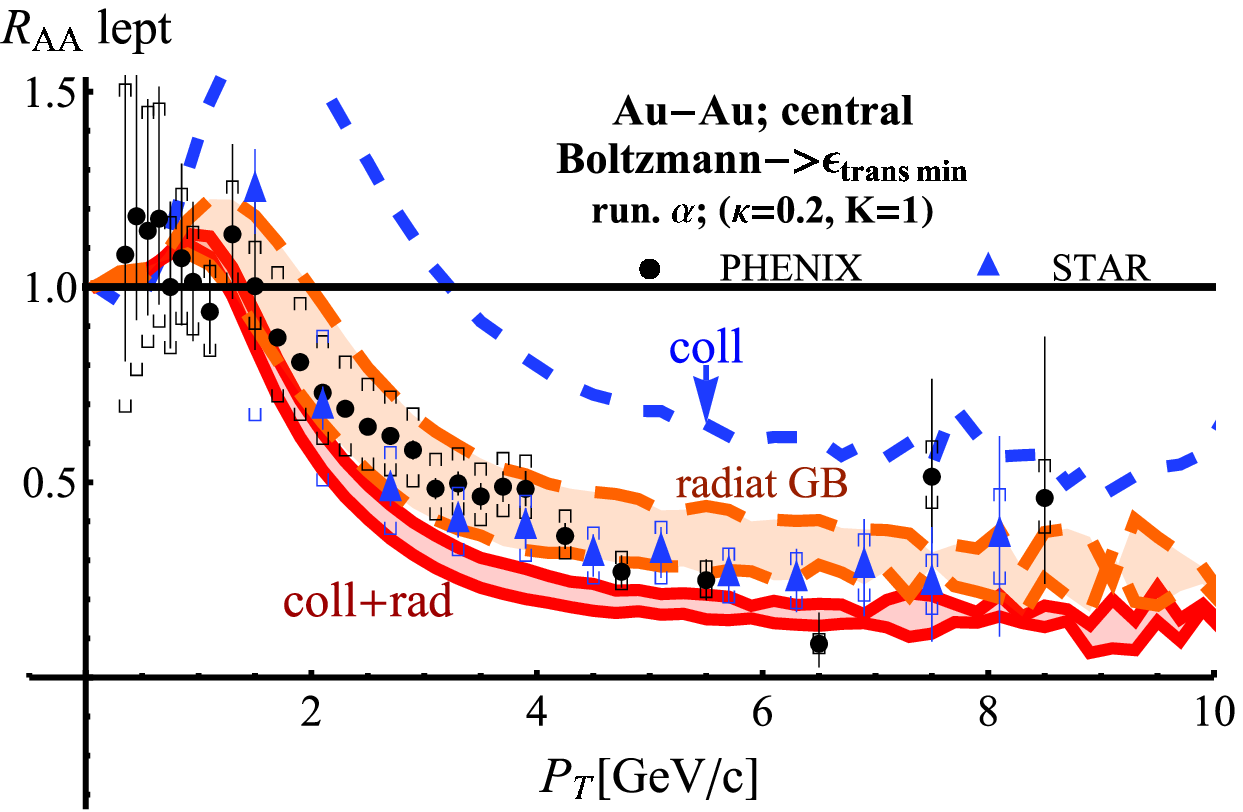,width=0.4\textwidth}
\epsfig{file=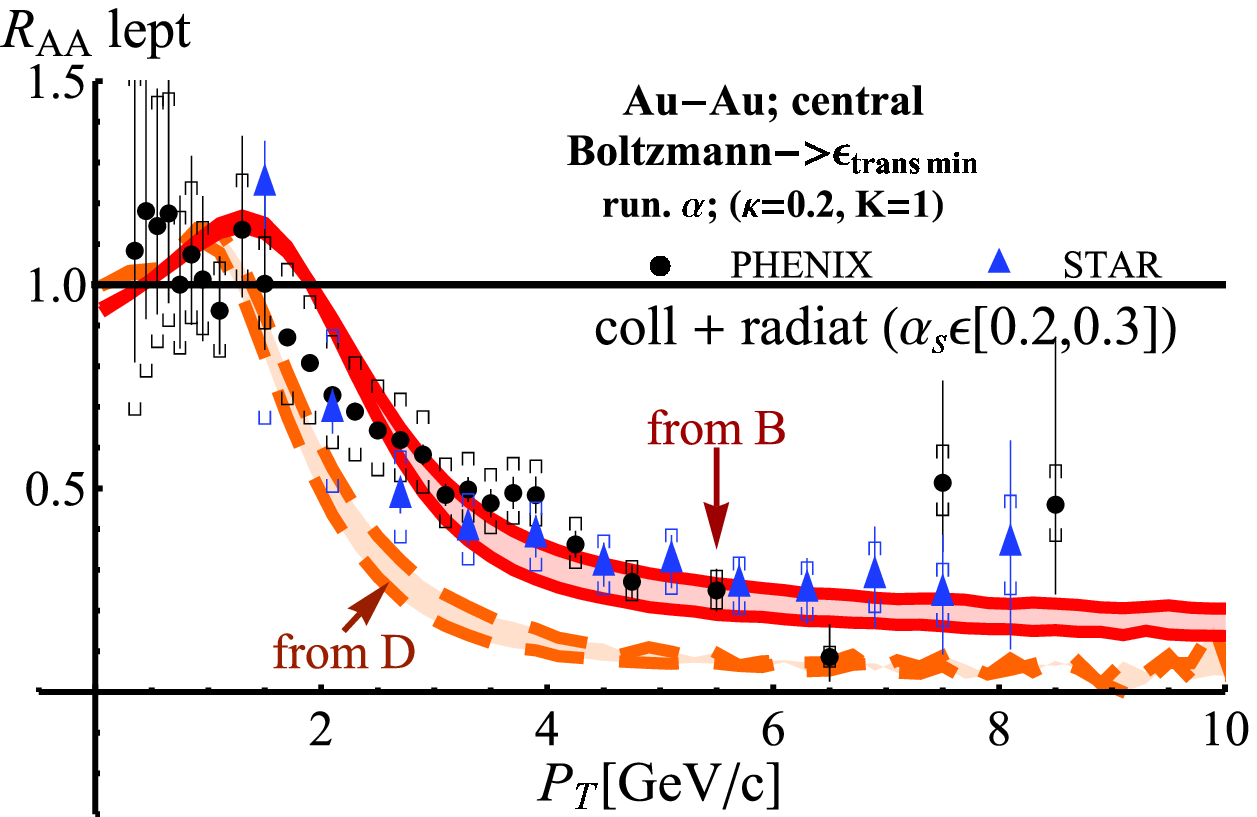,width=0.4\textwidth}
\epsfig{file=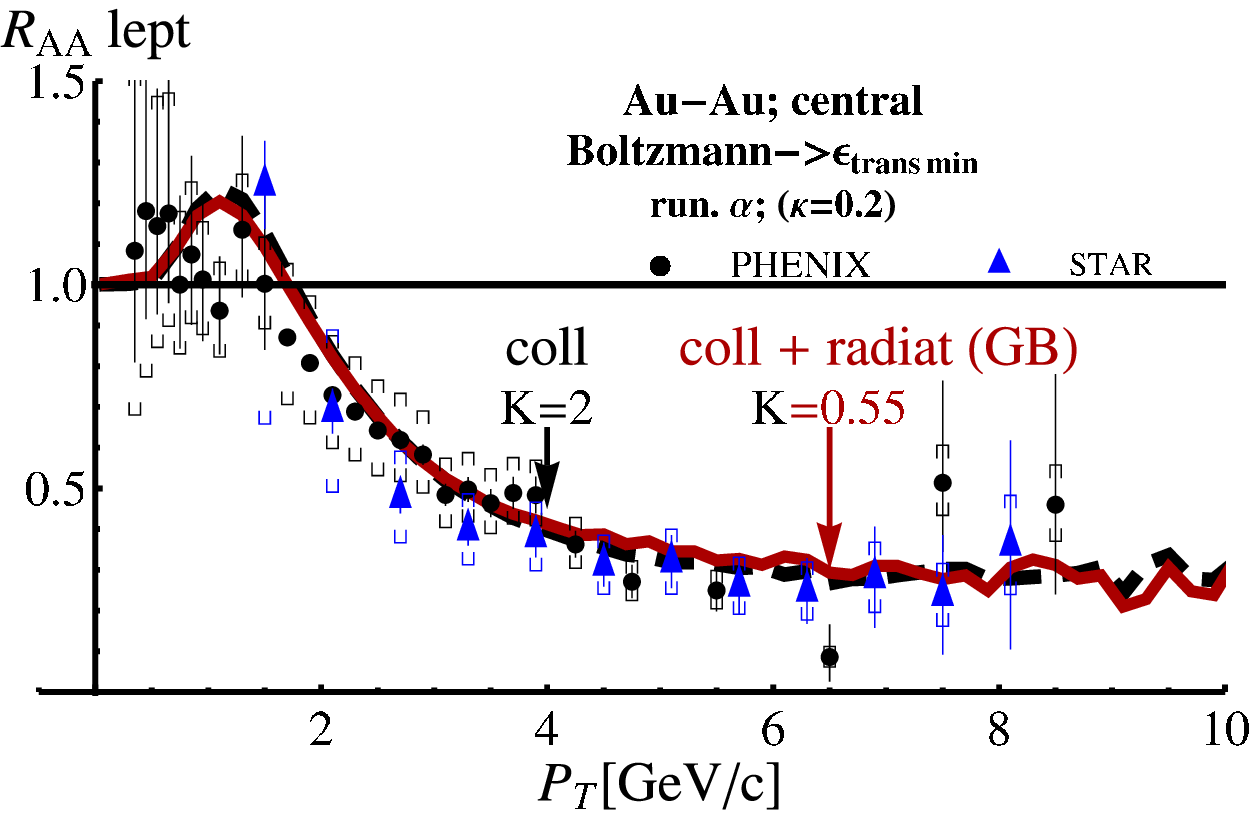,width=0.4\textwidth}
\epsfig{file=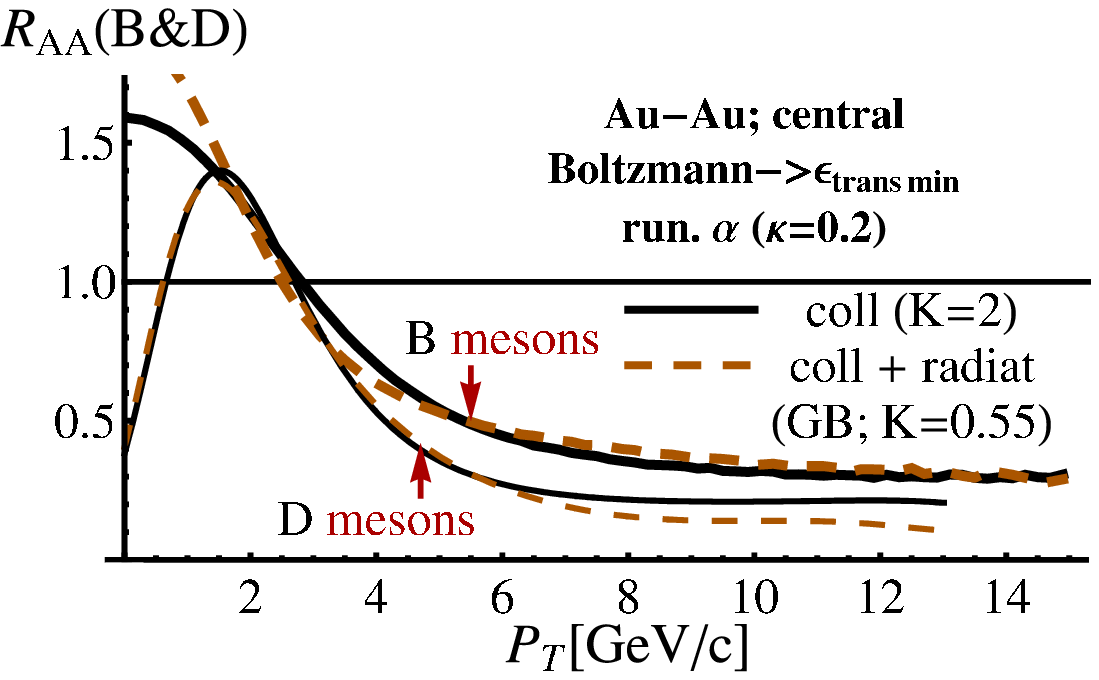,width=0.4\textwidth}
\end{center}
\caption{(Color online) top left: nuclear modification factor $R_{AA}$ of single leptons resulting
from the decay of heavy mesons (B and D) produced in central Au+Au ($\sqrt{s_{NN}}=200\,{\rm GeV}$), for 3 
different energy loss scenarios; top right: same for coll. + rad. case, separated for leptons resulting from 
$B$ and $D$ mesons decay; bottom left: same as top left in case of pure collisional and coll. + rad. mixture with 
respective multiplicative factors, $K=2$ and $K=0.55$, of the interaction rates; bottom right: same as bottom 
left for $B$ and $D$ mesons.}
\label{res1}
\end{figure}

Fig.\ref{res1} displays essential results in our approach. On the top-left panel, we show $R_{AA}$ as a function of 
$p_T$ in comparison with experimental Au-Au data, with and without radiative energy loss. 
As already discussed in \cite{Gossiaux:2008jv}, without radiative energy loss $R_{AA}$ is about 0.5 for large $p_T$ 
values and therefore well above the data. Radiative energy loss alone (shown as the shaded area bounded by calculation 
for $\alpha_s=0.2$ and 0.3 for the gluon emission vertex) reproduces almost the observed values of $R_{AA}$, 
while our results are slightly below the data if we add both types of energy losses. These observations are in 
contradiction with the established literature \cite{Djordjevic:2006a,Wicks:07,Armesto:06}. The reasons for this are 
twofold: a) the running coupling approach presented in \cite{Gossiaux:2008jv} naturally leads to $\hat{q}$ values 
(see fig. 6 of \cite{Gossiaux:2009mk}) that exceed the fixed $\alpha_s$ pQCD prediction \cite{Baier:02} and b) we 
limit ourselves, in this first study, to incoherent GB radiation and do not incorporate finite length nor LPM-like 
effects (see discussion below) which obviously quench the radiation spectra. In our view, a) constitutes a global
improvement, while b) is a shortcoming of our present description of radiative energy loss.

On the right hand side of the top panel we display $R_{AA} (p_T)$ separately for leptons from 
$D$ and $B$ meson decay. Leptons from D mesons are practically not present anymore at large $p_T$ values. 
$c$-quarks are indeed stronger quenched than $b$-quarks and contribute less to high $p_T$ leptons due to their 
softer fragmentation function. 
On the left-bottom panel, we proceed along the strategy of \cite{Phenix:08} and optimize our multiplicative 
coefficient $K$ according to the most central class of Au-Au collisions. What is astonishing is the fact that the 
functional form of $R_{AA} (p_T)$ is very similar for ``radiative + collisional'' and for ``collisional only'' 
energy loss and that they both provide a good description\footnote{At small and intermediate $p_T$, this claim
should be taken with a grain of salt for the ``coll. + rad.'' scenario, as the transition amplitude was derived 
in the large energy limit.} of the data, although the quenching at high $p_T$ is often 
considered as dominated by radiative processes. Although not displayed 
here\footnote{see http://omnis.if.ufrj.br/$\sim$sqm09/presentations/P.B.\underline{ }Gossiaux.pdf for explicit 
plots.}, an equally good agreement of both rescaled models is found for all Au+Au centrality classes as 
well as for the elliptical flow $v_2$. In fact, a detailed analysis demonstrates that the $R_{AA}$ observable is 
mostly sensitive to a succession of moderate energy transfers for which both processes have similar probability
distribution of fractional energy loss up to a global factor; this explains why rescaling both types of 
energy loss scenarios leads to similar agreement for $R_{AA}(p_T)$ and $v_2$. In short terms, the physics of 
$R_{AA}$ at RHIC is dominated by the Fokker-Planck regime, even at the largest $p_T$. This explains why 
models lacking radiative energy loss like \cite{VHR:05} are able to reproduce the data as well. On the other hand 
this implies that {\em radiative and collisional energy loss are not distinguishable on the basis of present 
n.p.s.e. experimental data we have analyzed}. This is our main conclusion and the main scientific message that we 
wanted to deliver at SQM 2009. 

To provide some perspective for future experiments, we have shown, on the bottom-right panel, the $R_{AA}$
for $B$ and $D$ mesons for both rescaled scenario ($K=2$ for ``coll.'' and $K=0.55$ for ``coll + rad. (GB)'').
For pure collisional and $p_T$ around 10 GeV, the ratio of heavy mesons $R_{AA}$, $R_{cb}(p_T)=
R_{AA}^D(p_T)/R_{AA}^B(p_T)$ is found to be of the order of 0.6~\cite{Gossiaux:2009hr}. When radiative energy 
loss is included, one has $R_{cb}(10)\approx 0.4$, which agrees with the value of 0.45 found in \cite{Wicks:07} 
and is still twice as large as predicted in the AdS/CFT approach \cite{Horowitz:2008b}. Thus, those various 
models could be deciphered with the help of the high-$p_T$ $R_{cb}$ ratio, although it is questionable 
whether upgrades of RHIC experiments will be sufficient to provide the mandatory resolution power, especially 
for the $D$-mesons.

The Landau Pomeranchuck Migdal (LPM) and finite length effects, which are of essential importance for the 
radiation off light partons, turn out to be less important for the heavy quarks because the finite mass reduces 
the (effective) formation length, found to be $l_f\approx\frac{x(1-x)E}{m_g^2+x^2 m^2}$ for vacuum radiation. 
For small as well as for large $x$, $l_f$ is smaller than the mean free path $\lambda_Q$; only for the emission 
off a heavy quarks around $x\approx m_g/m$, the formation length exceeds substantially $\lambda_Q$, but we 
are nevertheless closer to the incoherent regime than for radiation off l.q., as demonstrated 
in~\cite{Djordjevic:04a}. This was the primary motivation for neglecting LPM and finite length effects in this 
first implementation of radiative processes in MC$@_s$HQ. Looking back,
finding the same typical values for $R_{cb}$ as in \cite{Wicks:07} -- where those effects {\em are} taken into 
account -- might be the indication that the main source of discrepancy relies in the $\hat{q}$ value, that is
in the overall interaction rates. Although our approach \cite{Gossiaux:2008jv} complemented with radiative 
energy-loss constitutes a way out of level-2 s.e.p., it is still incomplete in the sense that {\em we have not yet 
investigated its consequences on the quenching of light hadrons}, i.e. the level-2 s.e.p.

In conclusion, we have shown that the combination of radiative and collisional energy loss, both calculated in 
a running $\alpha_s$ pQCD-inspired model, allows to describe simultaneously the nuclear modification factor
of n.p.s.e. for all centrality classes as well as the elliptical flow observed in Au-Au collisions at 
$\sqrt{s}=200$ AGeV. We have argued that radiative and collisional energy loss are not distinguishable on the basis 
of these data only. The influence of some theoretical uncertainties \cite{Armesto:06,Wicks:07} and of different 
expansion scenarios\footnote{different initial conditions, different freeze out temperatures and different equations 
of states in the hydrodynamical calculations.} -- i.e. the generalization of \cite{Bass:09} for heavy-quark
observables -- remains to be seen, although we do not expect it will affect our main conclusion. A detailed 
incorporation of the coherence effects (LPM and finite-length) is also mandatory, although a preliminary 
implementation of LPM effect in our framework, to be presented in an upcoming publication, was found of mere 
importance for $c$-quarks at RHIC and negligible for $b$-quarks.

\ack This work was performed under the ANR research program ``hadrons @ LHC'' 
(grant ANR-08-BLAN-0093-02) and the PCRD7/I3-HP program TORIC.\\

\end{document}